\documentclass[fleqn,12pt,twoside]{article}
\usepackage{espcrc1}
\usepackage{epsfig}
\usepackage{amsmath}
\usepackage{amssymb}
\usepackage{amsbsy}
\usepackage{amsfonts}
\usepackage{graphics}
\usepackage{latexsym}
\usepackage{epsf}
\usepackage{graphicx}

\newcommand{\bc}{\begin{center}}
\newcommand{\ec}{\end{center}}
\newcommand{\be}{\begin{equation}}
\newcommand{\ee}{\end{equation}}
\newcommand{\bea}{\begin{eqnarray}}
\newcommand{\eea}{\end{eqnarray}}
\newcommand{\bi}{\begin{itemize}}
\newcommand{\ei}{\end{itemize}}
\newcommand{\bt}{\begin{tabular}}
\newcommand{\et}{\end{tabular}}

\def\Nf{N_{\rm f}}

\def\mps{m_{\rm PS}}

\def\msbar{\overline{\rm MS}}

\def\ks{\kappa_{\rm sea}}
\def\kv{\kappa_{\rm val}}

\newcommand{\Dlr}{\stackrel{\leftrightarrow}{D}}

\title{
\begin{flushleft}
\vspace*{-2cm}
{\normalsize DESY 05-009}\\[-0.3em]
{\normalsize Edinburgh 2005/01}\\[-0.3em]
{\normalsize LU-ITP 2005/006}
\vspace*{0.15cm}
\end{flushleft}
Generalized parton distributions and transversity from 
full lattice QCD%
\thanks{Talks presented by Ph. H\"agler and J. Zanotti at Baryons 2004.}
}

\author{M. G\"ockeler\address{Institut f\"ur Theoretische Physik, 
Universit\"at Leipzig, D-04109 Leipzig, Germany}$^,$\address{
Institut f\"ur Theoretische Physik, Universit\"at Regensburg, 
D-93040 Regensburg, Germany},
Ph. H\"agler\address{Department of Physics and Astronomy, Vrije Universiteit,
1081 HV Amsterdam, NL},
R. Horsley\address{School of Physics, University of Edinburgh, 
Edinburgh EH9 3JZ, UK},
D. Pleiter\address{John von Neumann-Institut f\"ur Computing
NIC / DESY, 15738 Zeuthen, Germany},
P.E.L. Rakow\address{Theoretical Physics Division, Dep.~of 
Math.~Sciences, University of Liverpool, Liverpool L69 3BX, UK},
A. Sch\"afer$^{\rm b}$,
G. Schierholz$^{\rm e,}$\address{Deutsches Elektronen-Synchrotron  
DESY, 22603 Hamburg, Germany} and
J.M. Zanotti$^{\rm e}$
\emph{(QCDSF Collaboration)}}

\begin{document}

\maketitle

\begin{abstract}
We present here the latest results from the QCDSF
collaboration for moments of generalized parton distributions and
transversity in two-flavour QCD, including a preliminary analysis of
the pion mass dependence.
\end{abstract}

\vspace*{-2mm}
\section{INTRODUCTION}
\vspace*{-2mm}

For many years deep-inelastic scattering experiments have provided a
wealth of information regarding the quark and gluon content of the
nucleon, mainly through parton distribution functions which describe
the longitudinal momentum distributions of quarks and gluons in the
nucleon.  
Although the amount of knowledge gained from such experiments is
constantly increasing, little is known about the transverse
structure and angular momentum distribution of partons within the nucleon.
Generalized parton distributions (GPDs) \cite{GPD} have opened new
ways of studying the complex interplay of longitudinal momentum and
transverse coordinate space \cite{Diehl,Bu}, as well as spin and
orbital angular momentum degrees of freedom in the nucleon \cite{Ji}.
A full mapping of the parameter space spanned by GPDs
is an extremely extensive task which most probably needs support
from non-perturbative techniques like lattice simulations.

Besides the polarized and unpolarized GPDs, there exist also four
independent tensor/helicity flip GPDs $H_T(x,\xi,t), E_T, \widetilde
H_T$ and $\widetilde E_T$, as has been shown in \cite{Diehl:2001pm}.
The first one, $H_T(x,\xi,t)$, is called generalized transversity,
because it reproduces the transversity distribution in the foward
limit, $H_T(x,0,0)=\delta q(x)=h_1(x)$.
Since the quark tensor GPDs flip the helicity of the quarks, they do
not contribute to the deeply virtual Compton scattering (DVCS) process
$\gamma^* p\rightarrow \gamma p'$.
Although this could in principle be balanced by the production of a
transversely polarized vector meson, $\gamma^* p\rightarrow m_T p'$,
it has been shown that the corresponding amplitude vanishes at leading
twist to all orders in perturbation theory
\cite{Mankiewicz:1997uy,Diehl:1998pd,Collins:1999un}.
The only process we are currently aware of which gives access to the
generalized transversity is the diffractive double meson production
proposed in \cite{Ivanov:2002jj}.  It will be very interesting to see
if a measurement of tensor GPDs in this process is indeed feasible.
In any case, the situation seems to be much more difficult as compared
to the (un-)polarized GPDs, and an independent lattice calculation of
the moments of the helicity flip GPDs would be highly valuable.

There has been a large amount of activity within the lattice community
in the area of GPDs leading to some exciting results from the
QCDSF \cite{QCDSF-1,QCDSF-Cairns,Gockeler:2004vx,Gockeler:2004mn} and  
LHP \cite{MIT,MIT-2,Renner:2004ck,Hagler:2004er,Wolfram,Renner:2005sm}
collaborations.

Here we present the latest results from the QCDSF collaboration for
the first three moments of the unpolarized and polarized GPDs and
results for the tranversity GPDs. All results are preliminary.

\vspace*{-3mm}
\section{SIMULATION DETAILS}
\vspace*{-2mm}

We simulate with $\Nf=2$ dynamical configurations generated with
Wilson glue and non-perturbatively ${\cal O}(a)$ improved Wilson
fermions.
For five different values $\beta=5.20$, $5.25$, $5.26$,
$5.29$, $5.40$ and up to three different kappa values per beta
we have in collaboration with UKQCD generated ${\cal O}(2000-8000)$
trajectories.
Lattice spacings and spatial volumes vary between 0.075-0.123~fm and
(1.5-2.2~fm)$^3$ respectively.

Correlation functions are calculated on configurations
taken at a distance of 5-10 trajectories using 8-4 different locations of
the fermion source. We use binning to obtain an effective distance of
20 trajectories. The size of the bins has little effect on the
error, which indicates auto-correlations are small.
This work improves on previous calculations by adding one more sink
momentum, $\vec{p}_2$, and polarization, $\Gamma_1$. We use
$\vec{p}_0 = ( 0, 0, 0 )$,
$\vec{p}_1 = ( p, 0, 0 )$,
$\vec{p}_2  = ( 0, p, 0 )$
($p=2\pi/L_S$) and
$\Gamma_{\rm unpol} = \frac{1}{2}(1+\gamma_4)$,
$\Gamma_1 = \frac{1}{2}(1+\gamma_4)\, i\gamma_5\gamma_1$,
$\Gamma_2 = \frac{1}{2}(1+\gamma_4)\, i\gamma_5\gamma_2$.

\vspace*{-3mm}
\section{GENERALIZED PARTON DISTRIBUTIONS}
\vspace*{-2mm}

For a lattice calculation of GPDs, we work in Mellin-space
%where we use the operator product expansion (OPE)
to relate matrix
elements of local operators to Mellin moments of the GPDs. For quark
distributions, the twist-2 operators are
\vspace*{-5mm}
\bea
{\cal O}^{\{\mu_1\cdots\mu_n\}} &=& \overline{q} \, i \,
\gamma^{\{\mu_1}\, \Dlr^{\mu_2} \cdots
\Dlr^{\mu_n\}}\! q\ , \label{up-twist2} \\
\widetilde{{\cal O}}^{\{\mu_1\cdots\mu_n\}} &=& \overline{q} \, i \,
\gamma^5\gamma^{\{\mu_1}\, \Dlr^{\mu_2} \cdots
\Dlr^{\mu_n\}}\! q\ , \label{pol-twist2} \\
{\cal O}_T^{[\rho\{\mu_1]\cdots\mu_n\}} &=& \overline{q} \, i \,
\gamma^5\sigma^{\rho\{\mu_1}\gamma^{\mu_2}\, \Dlr^{\mu_3} \cdots
\Dlr^{\mu_n\}}\! q\ , \label{t-twist2} 
\eea
where $\Dlr = \frac{1}{2}(\overrightarrow{D} - \overleftarrow{D})$ and
$\{\cdots\}$ indicates symmetrization of indices and removal of
traces.
The non-forward matrix elements of the twist-2 operators
Eqs.~(\ref{up-twist2}, \ref{pol-twist2}, \ref{t-twist2}) specify the
$(n-1)^{th}$ moments of the spin-averaged, spin-dependent and
spin-flip generalized parton distributions, respectively.
In particular, for the unpolarized GPDs, we have
\vspace*{-3mm}
\bea
\!\!\int_{-1}^1 dx\, x^{n-1}\, H^q(x,\xi, \Delta^2) \!\!&=&\!\!
 H^q_{n}(\xi,\Delta^2)\ , \nonumber \\
\!\!\int_{-1}^1 dx\, x^{n-1}\, E^q(x,\xi, \Delta^2) \!\!&=&\!\!
 E^q_{n}(\xi,\Delta^2)\ ,
\eea
where \cite{Ji}
\vspace*{-6mm}
\bea
H^q_{n}(\xi,\Delta^2) &=& \sum_{i=0}^{\frac{n-1}{2}}
 A^q_{n,2i}(\Delta^2) (-2\xi)^{2i}
 + {\rm Mod}(n+1,2) C^q_{n}(\Delta^2) (-2\xi)^n \ ,\nonumber\\
E^q_{n}(\xi,\Delta^2) &=& \sum_{i=0}^{\frac{n-1}{2}}
 B^q_{n,2i}(\Delta^2) (-2\xi)^{2i}
 - {\rm Mod}(n+1,2) C^q_{n}(\Delta^2) (-2\xi)^n \ ,
\label{GPDmoments}
\eea
and the generalized form factors $A^q_{n,2i}(\Delta^2)$,
$B^q_{n,2i}(\Delta^2)$ and $C^q_{n}(\Delta^2)$ for the lowest three moments
are extracted from the nucleon matrix elements $\langle p'|{\cal
  O}^{\{\mu_1\cdots\mu_n\}}|p\rangle$ \cite{Ji}.
Note that the momentum transfer is given by $\Delta=p'-p$ with $t=\Delta^2$, 
while $\xi=-n\cdot \Delta/2$ denotes the longitudinal momentum transfer.
For the lowest moment, $A_{10}$ and $B_{10}$ are just the Dirac and
Pauli form factors $F_1$ and $F_2$, respectively. We also observe that
in the forward limit ($\Delta^2 = \xi = 0$), the moments of $H_q$
reduce to the moments of the unpolarized parton distribution $A_{n0}
= \langle x^{n-1}\rangle$.
Finally, the forward limit of the $x$-moment of the GPD $E$, $\int
dx\,x\,E(x,0,0) = B_{20}(0)$, allows for the determination of the
quark orbital angular momentum contribution to the nucleon spin,
$L^q=1/2(\langle x\rangle + B_{20} - \Delta q)$, where $\langle
x\rangle$ is the quark momentum fraction \cite{Ji}.

\begin{figure}[t]
%\vspace*{-2cm}
\hspace*{-3mm}
\includegraphics[height=9cm,angle=-90]{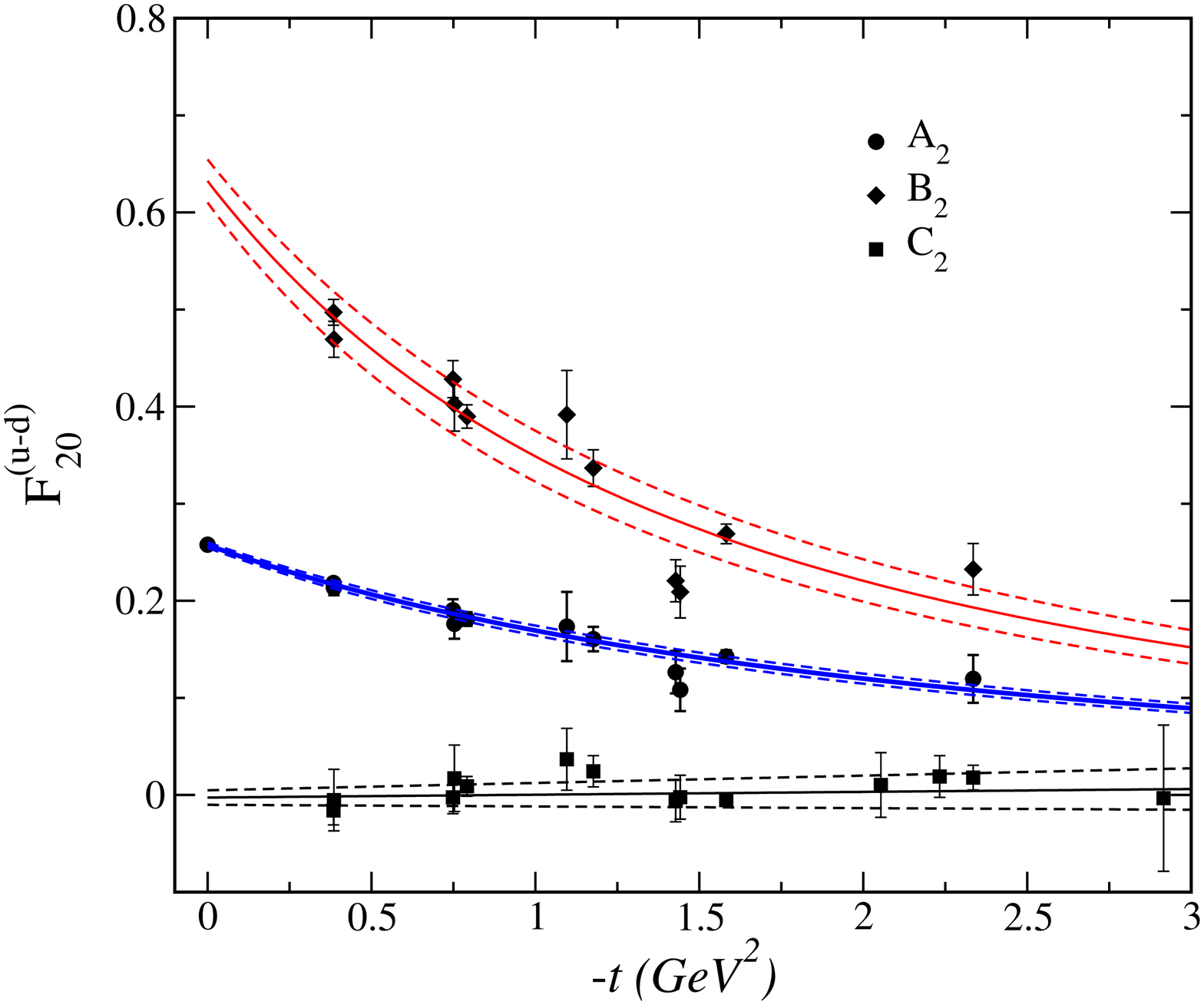}
\hspace*{-8mm}
\includegraphics[height=9cm,angle=-90]{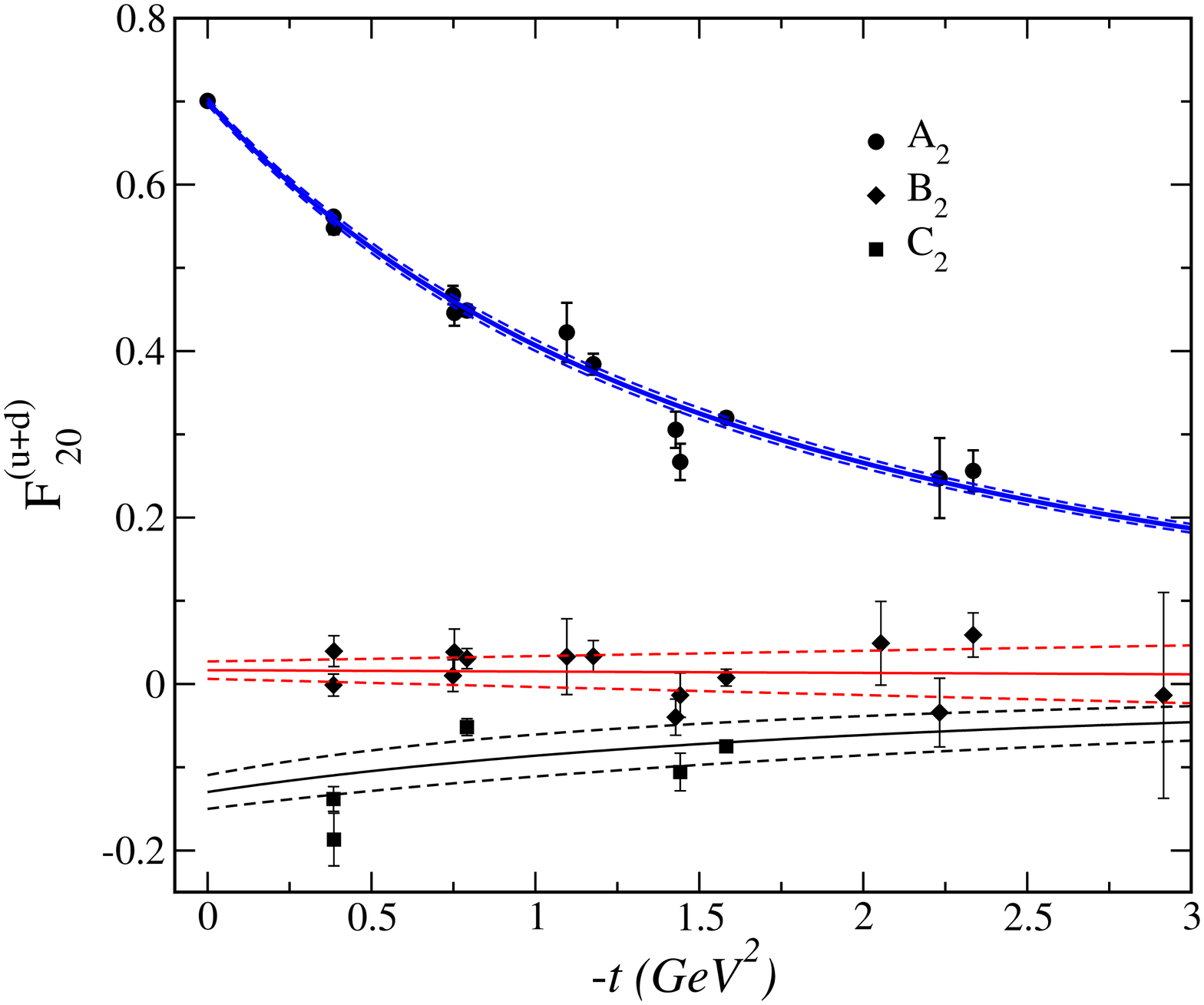}
\vspace*{-7mm}
\caption{\emph{Generalized form factors $A_{20},\ B_{20}$ and
  $C_{2}$, together with a dipole fit for the non-singlet, $u-d$
  (left), and singlet, $u+d$ (right).}}
\label{ABC20}
\vspace*{-7mm}
\end{figure}

In order to extract the non-forward matrix elements 
we compute ratios of three- and two-point functions
\vspace*{-5mm}
\bea
R(t,\tau;\vec{p}\,',\vec{p};{\cal O})\, &=&
 \frac{C_\Gamma (t,\tau;\vec{p}\,', \vec{p},{\cal O})} 
        {C_2(t,\vec{p}\,')}
 \left[
  \frac{C_2(\tau,\vec{p}\,') C_2(t,\vec{p}\,') C_2(t-\tau,\vec{p})}
  {C_2(\tau,\vec{p}) C_2(t,\vec{p}) C_2(t-\tau,\vec{p}\,')}
\right]^{\frac{1}{2} } \label{eq:ratio}\\
&\propto&\  \langle p'| {\cal O}_q|p\rangle \ , \nonumber
\eea
up to known kinematical factors as long as $0 \ll \tau \ll t \lesssim
\frac{1}{2} L_T$.

In Fig.~\ref{ABC20} we show, as an example, the generalized form
factors $A_{20},\ B_{20}$ and $C_{2}$ in the $\msbar$ scheme at
$\mu^2=4$ GeV$^2$ for the non-singlet, $u-d$ (left), and singlet, $u+d$
(right), on a $24^3\times48$ lattice at $\beta=5.40$ and
$\ks=\kv=0.13500$ corresponding to a lattice spacing, $a\,r_0 = 6.088$
and $m_\pi=\mps\approx970$ MeV.
We note here that in the calculation of the singlet matrix elements,
we neglect contributions coming from disconnected quark diagrams as
these are extremely computationally demanding.

\begin{figure}[t]
%\vspace*{-2cm}
\hspace*{-3mm}
\includegraphics[height=9cm,angle=-90]{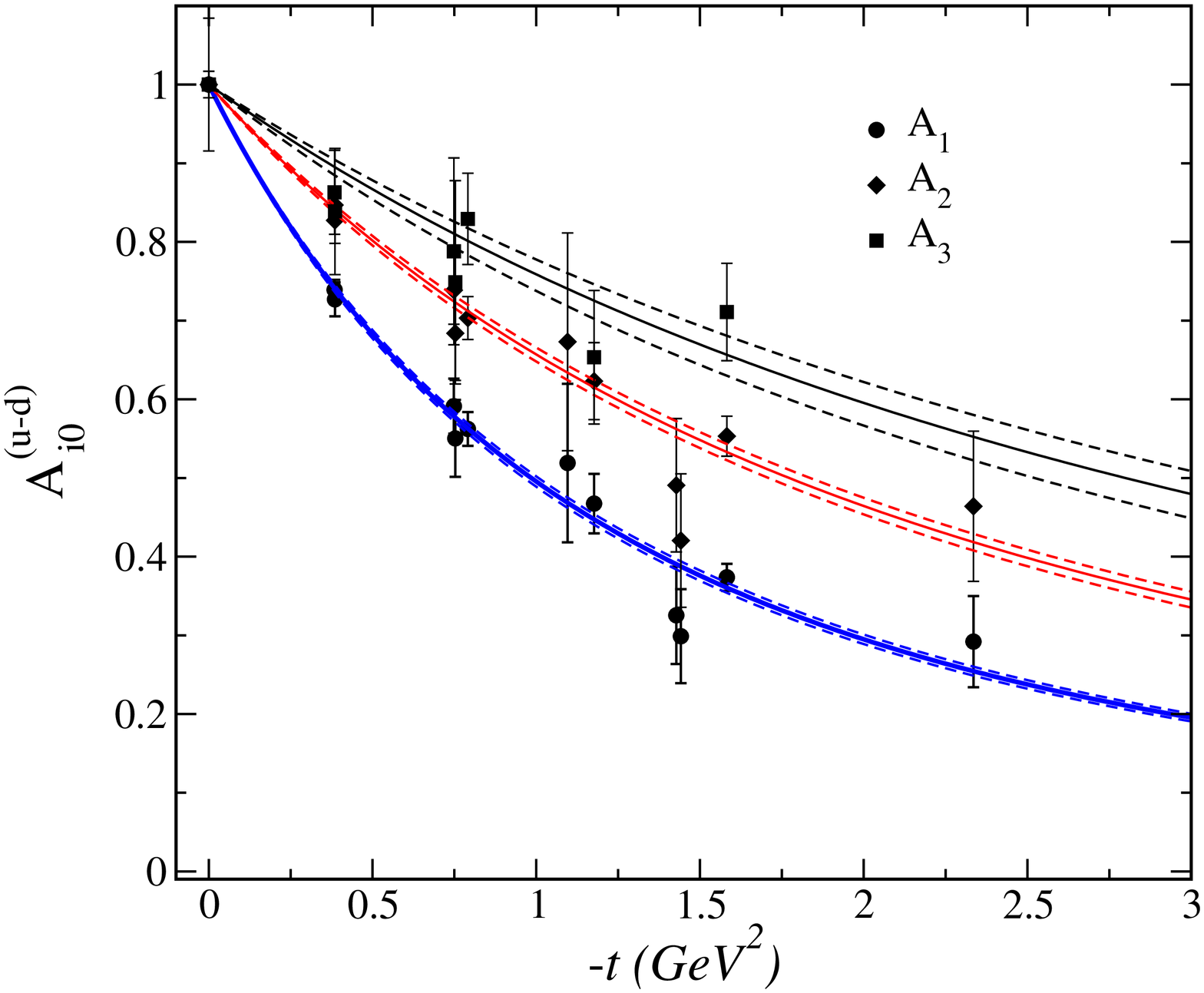}
\hspace*{-8mm}
\includegraphics[height=9cm,angle=-90]{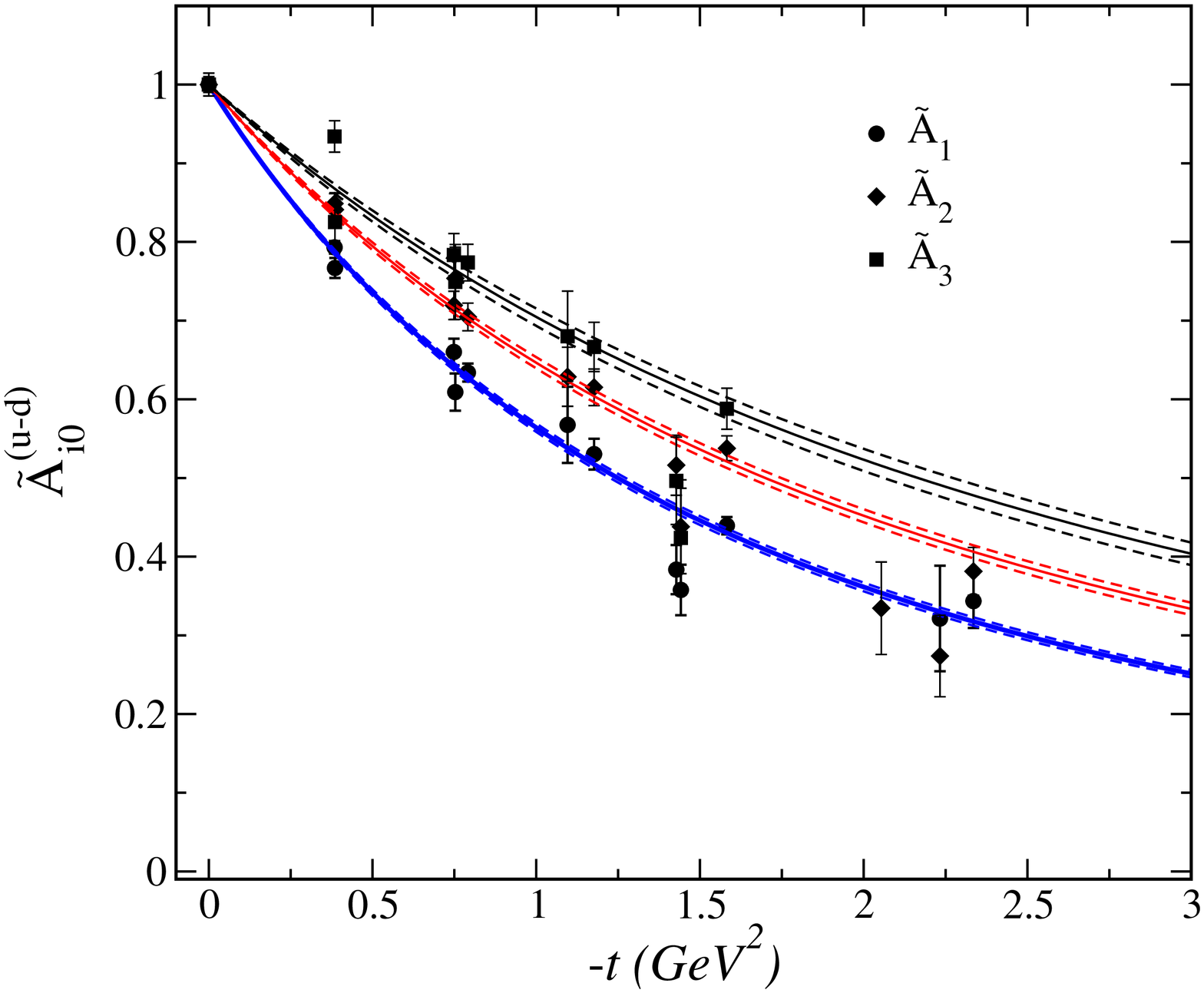}
\vspace*{-7mm}
\caption{\emph{Generalized form factors $A^{u-d}_{10},\ A^{u-d}_{20}$,
  $A^{u-d}_{30}$ (left), and $\widetilde{A}^{u-d}_{10},\
  \widetilde{A}^{u-d}_{20}$, $\widetilde{A}^{u-d}_{30}$ (right),
  together with a dipole fit. All form factors have been normalized to
  unity.}}
\label{A123}
\vspace*{-7mm}
\end{figure}

The generalized form factors $A^{u-d}_{20},\ A^{u+d}_{20}$ and
$B^{u-d}_{20}$ are well described
by the dipole ansatz
\vspace*{-3mm}
\be
A_n^q (\Delta^2) = \frac{A_n^q(0)}{\big( 1 - {\Delta^2/M_n^2}
  \big)^2} \ , 
\label{dipole}
\ee
while $C^{u-d}_{2}$ is consistent with zero. Our result for
$B^{u+d}_{2}$ reveals a small but positive value at this heavy quark
mass.  Future work will include a full non-perturbative analysis of
$\Delta q^{u+d}$ and $\Delta q^{u-d}$, together with a chiral
extrapolation, to reveal the quark contributions to the nucleon's total
spin and orbital angular momentum.

Burkardt~\cite{Bu} has shown that the spin-independent and
spin-dependent generalized parton distributions $H(x,0,\Delta^2)$ and
$\widetilde{H}(x,0,\Delta^2)$ gain a physical interpretation when
Fourier transformed to impact parameter space at
longitudinal momentum transfer $\xi=0$
\be
q(x,\vec{b}_\perp) =  \int \frac{d^2\Delta_\perp}{(2\pi)^2}\,
{\rm e}^{-i\vec{b}_\perp \cdot \vec{\Delta}_\perp}
H(x,0,-\Delta_\perp^2)\, ,
\label{fourier}
\ee
(and similar for the polarized $\Delta q(x,\vec{b}_\perp)$) where
$q(x,\vec{b}_\perp)$ is the probability density for a quark with
longitudinal momentum fraction $x$ and at transverse position (or
impact parameter) $\vec{b}_\perp$.

Burkardt \cite{Bu} also argued that $H(x,0,-\Delta_\perp^2)$ becomes
$\Delta_\perp^2$-independent as $x\rightarrow 1$ since, physically, we
expect the transverse size of the nucleon to decrease as $x$
increases,
i.e. $\lim_{x\rightarrow 1} q(x,\vec{b}_\perp) \propto
\delta^2(\vec{b}_\perp)$. As a result, we expect the slopes of the
moments of $H(x,0,-\Delta_\perp^2)$ in $\Delta_\perp^2$ to decrease as
we proceed to higher moments. 
This is also true for the polarized moments of
$\widetilde{H}(x,0,-\Delta_\perp^2)$, so from Eq.~(\ref{GPDmoments})
with $\xi=0$, we expect that the slopes of the generalized form
factors $A_{n0}(\Delta^2)$ and $\widetilde{A}_{n0}(\Delta^2)$ should
decrease with increasing $n$.

\begin{figure}[t]
\bc
\vspace*{-2cm}
\hspace*{-2cm}
\includegraphics[height=20cm,angle=-90]{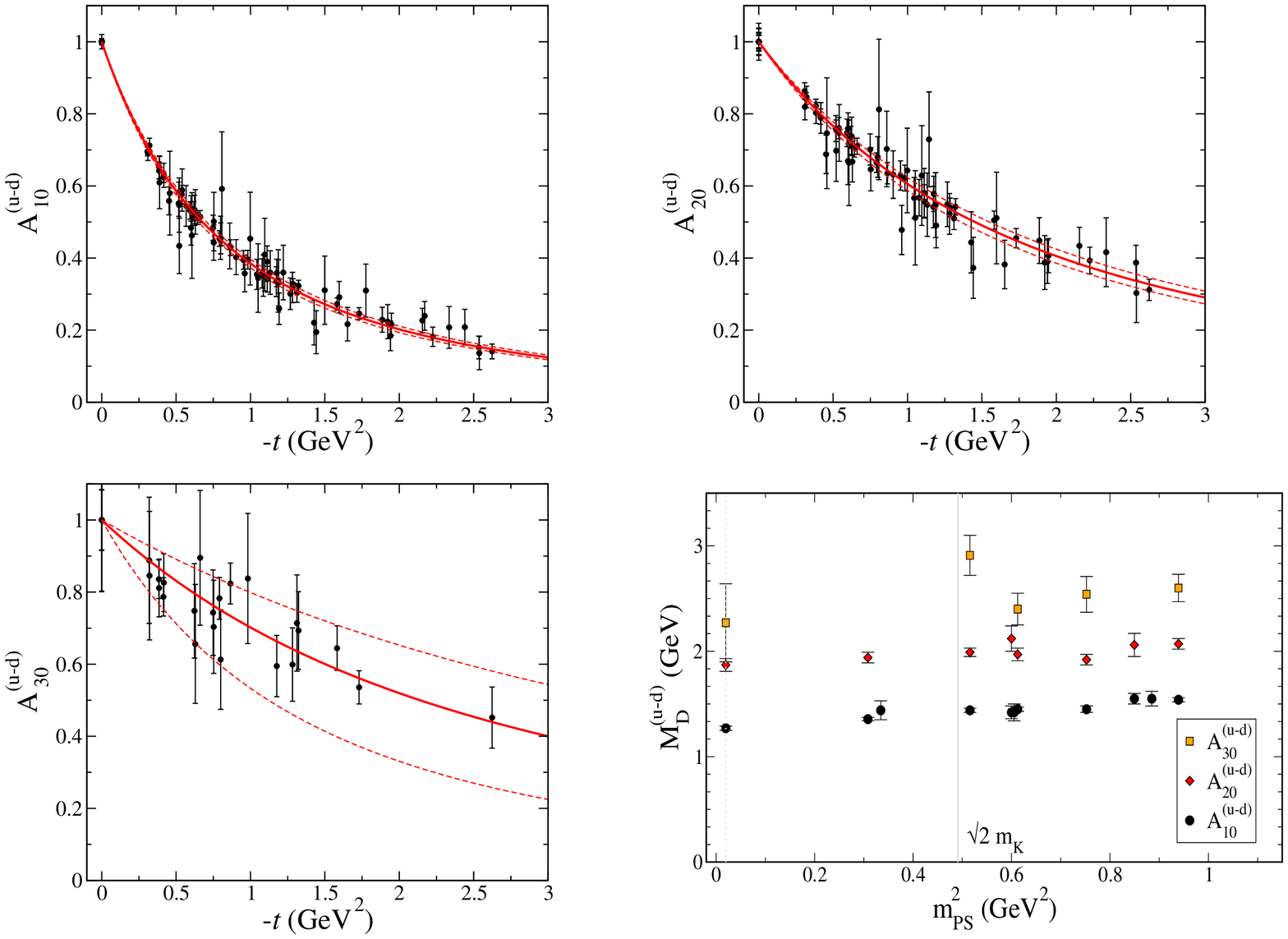}
\vspace*{-2.4cm}
\caption{\emph{Generalized form factors $A^{u-d}_{10},\ A^{u-d}_{20}$ and
  $A^{u-d}_{30}$, together with a dipole fit. All form factors have
  been normalized to unity. Dipole masses are extrapolated linearly to
the chiral limit.}}
\label{GlobalFits}
\ec
\vspace*{-12mm}
\end{figure}

In Fig.~\ref{A123}, we show the $\Delta^2$-dependence of
$A_{n0}(\Delta^2)$ (left) and $\widetilde{A}_{n0}(\Delta^2)$ (right),
$n=1,2,3$, for $\beta=5.40$, $\ks=\kv=0.13500$.
The form factors have been normalized to unity to make a comparison of
the slopes easier and as in Fig.~\ref{ABC20} we fit the form
factors with a dipole form as in Eq.~(\ref{dipole}).
We observe here that the form factors for the
unpolarized moments are well separated and that their
slopes do indeed decrease with increasing $n$ as predicted.
For the polarized moments, we observe a similar scenario, however here
the change in slope between the form factors is not as large.
The flattening of the GFFs $A_{n0}(\Delta^2)$ has first been observed
in Ref.~\cite{MIT-2}, where at the same time practically no change in
slope has been seen going from $\widetilde{A}_{20}(\Delta^2)$ to
$\widetilde{A}_{30}(\Delta^2)$.

Although fitting the form factors with a dipole is purely
phenomenological, (see Ref.~\cite{Diehl:2004cx} for an alternative ansatz)
it does provide us with a useful means to measure the
change in slope of the form factors by monitoring the extracted dipole
masses $(M_1,\, M_2,\, M_3)$ as we proceed to higher moments.
We have calculated these generalized form factors on a subset of our
full complement of $(\beta,\, \kappa)$ combinations and have extracted
the corresponding dipole masses. We plot these dipole masses in
the lower-right of Fig.~\ref{GlobalFits} as a function of $\mps^2$.
The values quoted in the chiral limit are calculated via a combined
dipole-mass/pion-mass fit 
\begin{equation}
\bar A_{n0}^{\text{dipole},m_\pi} (\Delta^2) = \frac{1}{\left( 1 - {t/(M_D^0+\alpha m_\pi^2)^2}
  \right)^2} \ ,
\label{chiraldipole}
\end{equation}
depending on the two parameters $M_D^0$ and $\alpha$, in order to
extract the maximum amount of information from our available
$(\beta,\, \kappa)$ combinations.
Having done the fit, we may shift the raw numbers to a common curve
given by Eq.(\ref{chiraldipole}) at the physical pion mass,
$m_{\pi,\text{phys}}=139$ MeV.  The results of this procedure are
shown in Fig.~\ref{GlobalFits} for the GFFs $A_{n0}$, $n=1,2,3$.

The important feature to note in Fig.~\ref{GlobalFits} is the
distinct separation between (and increase in magnitude of) the dipole
masses as we move to higher moments $(x\rightarrow 1)$.
Although the data available for $M_3$ is limited, the behaviour of all
dipole masses appears to be linear with $\mps^2$.
Consequently, we perform individual linear extrapolations of the
dipole masses $M_1,\, M_2,\, M_3$ to the physical pion mass, although
the findings of Ref.~\cite{Thomas} suggest that the chiral
extrapolation of the dipole masses of the electromagnetic form factors
may be non-linear.

\begin{figure}[t]
\bc
%\vspace*{-1cm}
\hspace*{-8mm}
\includegraphics[height=12cm,angle=-90]{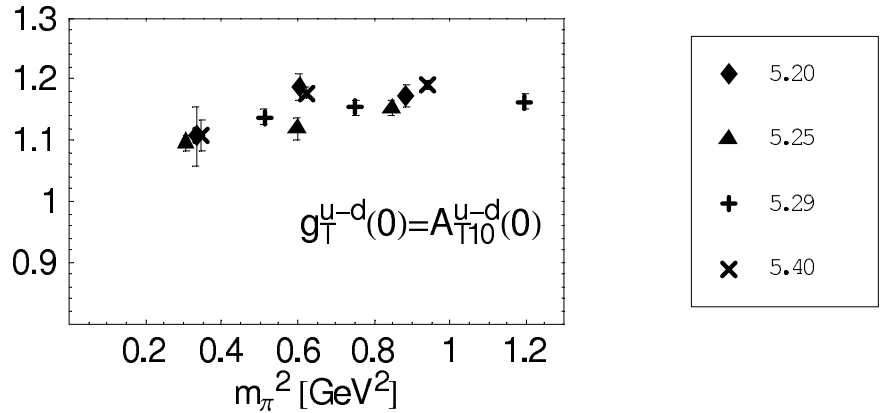}
\vspace*{-1cm}
\caption{\emph{The iso-vector tensor charge vs. $m_\pi^2$.}}
\label{AT10}
\ec
\vspace*{-10mm}
\end{figure}

If the dipole behavior Eq.~(\ref{dipole}) continues to hold for
the higher moments as well, and if we assume that the dipole masses
continue to grow in a Regge-like fashion, we may write
\be
\int_{-1}^1 \mbox{d}x\, x^{n-1} \,H_q(x,0,\Delta^2) = 
\frac{\langle x_q^{n-1}\rangle}{(1-\Delta^2/M_{n}^2)^2},
\label{H}
\ee
with $M_l^2 = \mbox{const.} + l/\alpha'$, where $\mbox{const.} \approx
-0.5$ GeV$^2$ and $1/\alpha' \approx 1.1$ GeV$^2$. 
This is sufficient to compute $H_q(x,0,\Delta^2)$ by means of an
inverse Mellin transform~\cite{DIS}.

Having done so, the desired probability distribution of finding a
parton of momentum fraction $x$ at the impact parameter
$\vec{b}_\perp$ can then be obtained by the Fourier transform of
Eq.~(\ref{fourier}).

%
%%%%%%%%%%%%%%%%%%%%%%%%%%%%%%%%%%%%%%%%%%%%%%%%%%%%%%%%%%%%%%%%%%%%%%%%%%%%%%%%%%%%%%%%%%%%%%%

We now turn our attention to our latest results for the first two
moments of the tensor GPDs.
Since the parameterizations for the Mellin-transformed tensor GPDs
have been derived recently in \cite{Hagler:2004yt,Chen:2004cg} in
terms of the GFFs $A_{Tni},\widetilde A_{Tni}, B_{Tni}$ and
$\widetilde B_{Tni}$, we are now in a position which allows for the
extraction of the moments of $H_T(x,\xi,t)$ etc. from our lattice
calculation.
The lowest moment of the generalized transversity, $\int dx
H_T(x,\xi,t) = A_{T10}(\Delta^2)$, is equal to the tensor form factor
$g_T(\Delta^2)=A_{T10}(\Delta^2)$.
Our results for the iso-vector tensor charge, $g^3_T(0)$, are shown in
Fig.~\ref{AT10} as a function of the pion mass squared.
The lattice results for $g^3_T$ have been non-perturbatively
renormalized and transformed to the $\msbar$ scheme at
$Q^2=4$ GeV$^2$.
The plot indicates that sizeable discretization and/or finite volume
effects may be present which we will study in detail in the near
future. A linear extrapolation of the isovector tensor charge to the
physical pion mass gives $g^3_T(0)=1.09 \pm 0.02$. 
A more advanced form for the chiral extrapolation of $g^3_T$
including effects from the pion cloud is also currently under
investigation \cite{Khan:2004vw,QCDSF-prep}.

\begin{figure}[t]
%\vspace*{-2cm}
\begin{center}
\hspace*{-3mm}
\includegraphics[height=17cm,angle=-90]{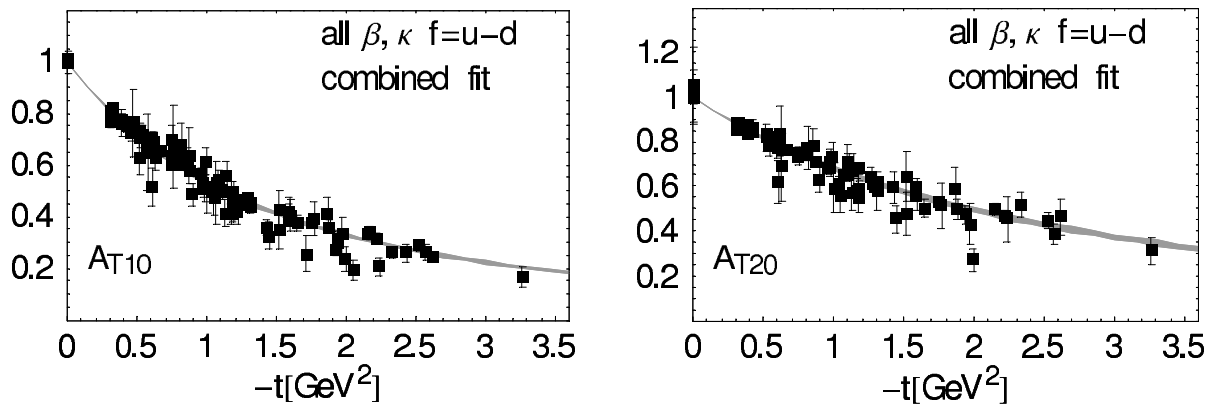}
\vspace*{-8mm}
\caption{\emph{Combined dipole-/pion-mass fits to
 the generalized form factors $A_{T10}$ and
  $A_{T20}$.}}
\label{ATi0}
\end{center}
\vspace*{-8mm}
\end{figure}

Fig.~\ref{ATi0} shows combined dipole-/pion-mass-fits of
$g^3_T(\Delta^2)$ and the $n=2$ -moment of the generalized
transversity, $A_{T20}(\Delta^2)$, following Eq.(\ref{chiraldipole}).
The corresponding dipole masses and fit-parameters are given by
\begin{eqnarray}
A_{T10}\, &:& \, M_D^0 = 1.63\!\pm\! .02 \text{ GeV}, \,\,\, \alpha= 0.10\!\pm\! .03 \text{ GeV}^{-2},
\,\,\, \chi^2/\text{DOF}=1.1\nonumber\\
A_{T20}\, &:& \, M_D^0 = 2.17\!\pm\! .04 \text{ GeV}, \,\,\, \alpha= -0.06\!\pm\! .06 \text{ GeV}^{-2},
\,\,\, \chi^2/\text{DOF}=0.9,
\label{FitRes}
\end{eqnarray}
showing an increase in the dipole mass, going from the $n=1$ to the
$n=2$-Mellin-moment.  The errors in Eq. (\ref{FitRes}) are purely
statistical. Results for the other tensor GFFs $\widetilde
A_{Tn0}, B_{Tn0}$ and $\widetilde B_{Tn0}$ for $n=1,2$ will be
presented in a future publication.  For the tensor GPDs, the number
$N$ of GFFs grows rapidly with $n$, i.e. $N=2\lceil n/2 \rceil+n$.
For $n=3$ we would have to extract simultaneously seven independent GFFs from the lattice,
which poses quite a challenge.
We plan to study the relevant two-derivative tensor
operators on the lattice to see if and to what extent an extraction is possible.

\vspace*{-3mm}
\section*{ACKNOWLEDGEMENTS}
\vspace*{-2mm}

The numerical calculations have been performed on the Hitachi SR8000
at LRZ (Munich), on the Cray T3E at EPCC (Edinburgh) under PPARC grant
PPA/G/S/1998/00777 \cite{UKQCD}, and on the APEmille at NIC/DESY
(Zeuthen). This work is supported in part by the DFG, by the EU
Integrated Infrastructure Initiative Hadron Physics under contract
number RII3-CT-2004-506078 and by the Helmholtz Assciation, contract
number VH-NG-004.  We thank A.~Irving for providing $r_0/a$ prior to
publication.


\vspace*{-3mm}
\begin{thebibliography}{99}

\bibitem{GPD} D. M\"uller {\it et al.}, Fortsch. Phys. \textbf{42}, 101 (1994);
  A.V. Radyushkin, Phys. Rev. D \textbf{56}, 5524
  (1997); M. Diehl {\it et al.}, Phys. Lett. B \textbf{411}, 193 (1997).

\bibitem{Diehl} M. Diehl, Eur. Phys. J. C \textbf{25}, 223 (2002). 

\bibitem{Bu} M. Burkardt, Int. J. Mod. Phys. A \textbf{18}, 173 (2003).

\bibitem{Ji} X. Ji, Phys. Rev. Lett. \textbf{78}, 610 (1997);
             X. Ji, J. Phys. G \textbf{24}, 1181 (1998).

\bibitem{Diehl:2001pm}  M.~Diehl,
%``Generalized parton distributions with helicity flip,''
Eur.\ Phys.\ J.\ C \textbf{19} (2001) 485.
%%CITATION = HEP-PH 0101335;%%

%\cite{Mankiewicz:1997uy}
\bibitem{Mankiewicz:1997uy}
L.~Mankiewicz, G.~Piller and T.~Weigl,
%``Hard exclusive meson production and nonforward parton distributions,''
Eur.\ Phys.\ J.\ C {\bf 5} (1998) 119.
%{\tt hep-ph/9711227}.
%%CITATION = HEP-PH 9711227;%%

%\cite{Diehl:1998pd}
\bibitem{Diehl:1998pd}
M.~Diehl, T.~Gousset and B.~Pire,
%``Exclusive electroproduction of vector mesons and transversity
%distributions,''
Phys.\ Rev.\ D {\bf 59} (1999) 034023.
%{\tt hep-ph/9808479}.
%%CITATION = HEP-PH 9808479;%%

%\cite{Collins:1999un}
\bibitem{Collins:1999un}
J.~C.~Collins and M.~Diehl,
%``Transversity distribution does not contribute to hard exclusive
%electroproduction of mesons,''
Phys.\ Rev.\ D {\bf 61} (2000) 114015.
%{\tt hep-ph/9907498}.
%%CITATION = HEP-PH 9907498;%%

%\cite{Ivanov:2002jj}
\bibitem{Ivanov:2002jj}
D.~Y.~Ivanov {\it et al.}, 
%B.~Pire, L.~Szymanowski and O.~V.~Teryaev,
%``Probing chiral-odd GPD's in diffractive electroproduction of two vector
%mesons,''
Phys.\ Lett.\ B {\bf 550} (2002) 65.
%{\tt hep-ph/0209300}.
%%CITATION = HEP-PH 0209300;%%

\bibitem{QCDSF-1} M. G\"ockeler {\it et al.}, Phys.\ Rev.\ Lett.\  \textbf{92},
042002 (2004).
%{\tt hep-ph/0304249}.
%%CITATION = HEP-PH 0304249;%%
%\cite{Gockeler:2004vx}

\bibitem{QCDSF-Cairns} T.~Bakeyev {\it et al.}, 
Nucl.\ Phys.\ Proc.\ Suppl.\  {\bf 128}, 82 (2004).
%{\tt hep-lat/0311017}.
%%CITATION = HEP-LAT 0311017;%%

\bibitem{Gockeler:2004vx}
M.~G\"ockeler {\it et al.},
%``Generalized parton distributions and structure functions from full lattice
%QCD,''
{\tt hep-lat/0409162}.
%%CITATION = HEP-LAT 0409162;%%

%\cite{Gockeler:2004mn}
\bibitem{Gockeler:2004mn}
M.~G\"ockeler {\it et al.},
%``Generalized parton distributions in full lattice QCD,''
{\tt hep-lat/0410023}.
%%CITATION = HEP-LAT 0410023;%%

\bibitem{MIT} Ph. H\"agler {\it et al.}, Phys. Rev. D \textbf{68} (2003) 034505.

\bibitem{MIT-2}
Ph.~H\"agler {\it et al.}, 
%J.~W.~Negele, D.~B.~Renner, W.~Schroers, T.~Lippert and K.~Schilling
%[LHPC],
%``Transverse structure of nucleon parton distributions from lattice QCD,''
Phys.\ Rev.\ Lett.\ \textbf{93} (2004) 112001.
%%CITATION = HEP-LAT 0312014;%%

%\cite{Renner:2004ck}
\bibitem{Renner:2004ck}
D.~B.~Renner {\it et al.},
%``Hadronic physics with domain-wall valence and improved staggered sea
%quarks,''
{\tt hep-lat/0409130}.
%%CITATION = HEP-LAT 0409130;%%

%\cite{Hagler:2004er}
\bibitem{Hagler:2004er}
Ph.~H\"agler {\it et al.}, 
%``Helicity dependent and independent generalized parton distributions of the
%nucleon in lattice QCD,''
{\tt hep-ph/0410017}.
%%CITATION = HEP-PH 0410017;%%

\bibitem{Wolfram} Talk by W. Schroers, these proceedings.

%\cite{Renner:2005sm}
\bibitem{Renner:2005sm}
D.~B.~Renner,
%``Generalized parton distributions from lattice QCD,''
{\tt hep-lat/0501005}.
%%CITATION = HEP-LAT 0501005;%%

%\cite{Diehl:2004cx}
\bibitem{Diehl:2004cx}
M.~Diehl, T.~Feldmann, R.~Jakob and P.~Kroll,
%``Generalized parton distributions from nucleon form factor data,''
{\tt hep-ph/0408173}.
%%CITATION = HEP-PH 0408173;%%

\bibitem{Thomas} J.D. Ashley {\it et al.}, Eur.\ Phys.\ J.\ A {\bf 19}, 9 (2004).
%{\tt hep-lat/0308024}.
%%CITATION = HEP-LAT 0308024;%%

\bibitem{DIS} M. G\"ockeler {\it et al.}, {\tt hep-ph/9711245}.

%\cite{Hagler:2004yt}
\bibitem{Hagler:2004yt}
Ph.~H\"agler,
%``Form-factor decomposition of generalized parton distributions at leading
%twist,''
Phys.\ Lett.\ B {\bf 594} (2004) 164. 
%{\tt hep-ph/0404138}.
%%CITATION = HEP-PH 0404138;%%

%\cite{Chen:2004cg}
\bibitem{Chen:2004cg}
Z.~Chen and X.~D.~Ji,
%``Counting and tensorial properties of twist-two helicity-flip nucleon form
%factors,''
{\tt hep-ph/0404276}.
%%CITATION = HEP-PH 0404276;%%

%\cite{Khan:2004vw}
\bibitem{Khan:2004vw}
A.~A.~Khan {\it et al.},
%``Axial and tensor charge of the nucleon with dynamical fermions,''
{\tt hep-lat/0409161}.
%%CITATION = HEP-LAT 0409161;%%

\bibitem{QCDSF-prep} M. G\"ockeler {\it et al.}, in preparation.

\bibitem{UKQCD} C.~R. Allton {\it et al.},
                Phys. Rev. D \textbf{65} (2002) 054502.

\end{thebibliography}
\end{document}